\documentstyle[aps,pre]{revtex}
\begin{document}

\draft

\title{RENORMALIZATION GROUP, OPERATOR PRODUCT EXPANSION AND ANOMALOUS
SCALING IN MODELS OF PASSIVE TURBULENT ADVECTION}
\author{L.Ts.~Adzhemyan, N.V.~Antonov and A.N.~Vasil'ev}
\address{Department of Theoretical Physics, St.~Petersburg University,
Uljanovskaja 1, St.~Petersburg, Petrodvorez, 198504 Russia}

\maketitle
\begin{abstract}
The field theoretic renormalization group is applied to Kraichnan's
model of a passive scalar quantity advected by the Gaussian velocity field
with the pair correlation function $\propto\delta(t-t')/k^{d+\varepsilon}$.
Inertial-range anomalous scaling for the structure functions and various
pair correlators is established as a consequence of the existence in the
corresponding operator product expansions of ``dangerous'' composite
opera\-tors (powers of the local dissipation rate), whose {\it negative}
critical dimensions determine anomalous exponents. The latter are
calculated to order $\varepsilon^{3}$ of the $\varepsilon$ expansion
(three-loop approximation). Submitted to {\bf Acta Physica Slovaca}.
\end{abstract}

In recent years, considerable progress has been achieved in the understanding
of intermittency and anomalous scaling of fluid turbulence. The crucial role
in these studies was played by a simple model of a passive scalar quantity
advected by a random Gaussian field, white in time and self-similar in space,
the so-called Kraichnan's rapid-change model \cite{Kraich1}. There, for the
first time the existence of anomalous scaling was established on the basis of
a microscopic model \cite{Kraich2} and the corresponding anomalous exponents
were calculated within controlled approximations [3--6]
and a systematic perturbation expansion in a formal small parameter
\cite{RG}. Detailed review of the recent theoretical research on the
passive scalar problem and the bibliography can be found in \cite{FGV}.

The advection of a passive scalar field $\theta(x)\equiv\theta(t,{\bf x})$
is described by the stochastic equation
\begin{equation}
\partial _t\theta+ (v_{i}\partial_{i}) \theta =\nu _0\Delta \theta+f,
\label{1}
\end{equation}
where $\partial _t \equiv \partial /\partial t$,
$\partial _i \equiv \partial /\partial x_{i}$, $\nu _0$
is the molecular diffusivity coefficient, $\Delta$ is the Laplace operator,
${\bf v}(x)\equiv \{v_{i}(x)\}$ is the transverse (owing to the
incompressibility) velocity field, and $f\equiv f(x)$ is an artificial
Gaussian scalar noise with zero mean and correlator
\begin{equation}
\langle  f(x)  f(x')\rangle = \delta(t-t')\, C(r/L), \quad
r=|{\bf x}-{\bf x'}|.
\label{2}
\end{equation}
The parameter $L$ is an integral scale related to the scalar noise and
$C(r/L)$ is some function finite as $L\to\infty$. Without loss of
generality, one can take $L=\infty$ and $C(0)=1$.

In the real problem, the field  ${\bf v}(x)$ satisfies the Navier--Stokes
equation. In the rapid-change model it obeys a Gaussian distribution with
zero mean and correlator
\begin{eqnarray}
\langle v_{i}(x) v_{j}(x')\rangle = D_{0}\,
\frac{\delta(t-t')}{(2\pi)^d}
\int d{\bf k}\, P_{ij}({\bf k})\,
k^{-d-\varepsilon}\, \exp [{\rm i}{\bf k}\cdot({\bf x}-{\bf x'})] ,
\label{3}
\end{eqnarray}
where $P_{ij}({\bf k}) = \delta _{ij} - k_i k_j / k^2$ is the transverse
projector, $k\equiv |{\bf k}|$, $D_{0}>0$ is an amplitude factor, $d$ is the
dimensionality of the ${\bf x}$ space and $0<\varepsilon<2$ is a parameter
with the real (``Kolmogorov'') value $\varepsilon=4/3$. The infrared (IR)
regularization is provided by the cut-off in the integral (\ref{3}) from
below at $k\simeq m$, where $m\equiv 1/\ell$ is the reciprocal of another
integral scale $\ell$; the precise form of the cut-off is unessential.
The relation $D_{0}/\nu_0 = \Lambda^{\varepsilon}$ defines the
characteristic ultraviolet (UV) momentum scale $\Lambda$.

The issue of interest is, in particular, the behaviour of the equal-time
structure functions $S_{n}(r) =\big\langle[\theta(t,{\bf x})-
\theta(t,{\bf x'})]^{n}\big\rangle$,  $r =|{\bf x}-{\bf x'}|$,
in the inertial range $\Lambda \gg 1/r \gg m$. (With our assumptions,
the odd structure functions vanish; they become nontrivial if the
correlation function $\langle vf \rangle$ is nonzero or a constant
gradient of the scalar field is imposed.) If we neglect the advection,
that is, the nonlinearity in (\ref{1}), these functions can be easily
calculated: $S_{n}(r) = C_{n} \nu_0^{-n}r^{2n}$ with some constants
$C_{n}$. In the full nonlinear problem, they become dependent on two
additional variables $\Lambda r$ and $mr$:
\begin{equation}
S_{n}(r)= \nu_0^{-n}r^{2n} F_{n}(\Lambda r, mr).
\label{4}
\end{equation}
The analysis by Refs. \cite{GK,Falk1} shows that the behaviour of the
functions $F_{n}$ in (\ref{4}) at $\Lambda r\to\infty$, $mr\to0$
(inertial range) is nontrivial:
\begin{equation}
S_{n}(r)\simeq C_{n}\nu_0^{-n}r^{2n} (\Lambda r)^{-\alpha_{n}}
(mr)^{-\beta_{n}}  ,
\label{5}
\end{equation}
that is, the dependence on the both scales persists. Moreover, the exponent
$\beta_{n}$ is a nonlinear function of $n$, the phenomenon referred to as
``anomalous scaling'' in the theory of turbulence.

Relations (\ref{5}) and explicit expressions for the exponents $\alpha_{n}$
and $\beta_{n}$ up to the first order in $1/d$ and $\varepsilon$ were derived
in \cite{GK,Falk1}. Within the ``zero-mode approach'' to Kraichnan's model,
developed in those papers, anomalous exponents are related to the zero
modes (unforced solutions) of the closed exact equations satisfied by the
equal-time correlations.

In Ref. \cite{RG} and subsequent papers [9--17], the
field theoretic renormalization group (RG) and operator product expansion
(OPE) were applied to the rapid-change model and its descendants. In the RG
approach, anomalous scaling emerges as a consequence of the existence in the
model of composite operators with negative scaling dimensions, identified
with the anomalous exponents. This allows one to construct a systematic
perturbation expansion for the anomalous exponents, analogous to the famous
$\varepsilon$ expansion in the RG theory of critical behaviour, and to
calculate the exponents to the second \cite{RG,RG1,RG2} and third
\cite{cube,kub} orders. For passively advected {\it vector} fields, any
calculation of the exponents for higher-order correlations calls for the RG
techniques already in the lowest-order approximation [15--17]. Besides the
calculational efficiency, an important advantage of the RG approach is its
relative universality: it is not related to the aforementioned
``solvability'' of the rapid-change model and can also be applied to the
case of finite correlation time or non-Gaussian advecting field \cite{RG3}.

Below we briefly review the main ideas and results of the RG approach to the
rapid-change model; detailed exposition and more references can be found in
\cite{RG,RG3,kub,vektor}.

The solution proceeds in two stages. In the first stage, the RG equation is
employed to find the asymptotic behaviour of the functions $F_{n}$ in
(\ref{4}) with respect to their first argument ($\Lambda r\to\infty$) at
fixed $mr$. This gives
\begin{equation}
S_{n}(r)\simeq \nu_0^{-n}r^{2n} (\Lambda r)^{-\alpha_{n}} F_{n}(mr).
\label{6}
\end{equation}
The total power of $r$, namely $2n-\alpha_{n}$, is (up to the minus sign)
the critical dimension of the quantity on the left-hand side. However,
the form of the function $F_{n}(mr)$ cannot be determined by the RG equation.

Expression (\ref{4}) is the special example of the general property
established by the RG method for problem (\ref{1})--(\ref{3}): the property
of the IR scale invariance. Similar asymptotic expressions can be obtained
for all correlation functions; in particular, for the equal-time pair
correlation functions $S_{nk}(r)=\langle \Phi_{n}(x) \Phi_{k}( x')\rangle$
of the monomials
$\Phi_{n}=[\partial_{i}\theta( x)\partial_{i}\theta( x)]^{n}$
one obtains:
\begin{equation}
S_{nk}(r)\simeq \nu_0^{-n-k} (\Lambda r)^{-\gamma_{nk}} F_{nk}(mr).
\label{7}
\end{equation}
In the RG technique, the exponents $\alpha_{n}$ and $\gamma_{nk}$
in (\ref{6}), (\ref{7}) are calculated in the form of series in the
parameter $\varepsilon$ from (\ref{2}), which is therefore the analogue
of the quantity $\varepsilon=4-d$ in the $\phi^{4}$ model of critical
behaviour.

From the general RG viewpoint, model (\ref{1})--(\ref{3}) is simpler than
the conventional $\phi^{4}$ model in a few respects: the coordinate of the
fixed point is exactly determined by the one-loop approximation,
renormalization of the primary field $\theta(x)$ and its powers $\theta^n(x)$
(``composite operators'' in the field theoretic language) is absent; the
``mass'' $m$ is not renormalized. Therefore the exponents $\alpha_{n}$ in
(\ref{6}) are found exactly, $\alpha_{n}=n\varepsilon$, that is, they have no
corrections of order $\varepsilon^{2}$, $\varepsilon^{3}$ and so on.
However the exponents $\gamma_{nk}$, given by the sums
$\gamma_{nk}=\Delta_{n}+\Delta_{k}$ of the critical dimensions
$\Delta_{n}$ of the composite operators $\Phi_{n}$, have nontrivial
$\varepsilon$ expansion. The knowledge of these dimensions allows one to
find the asymptotic form of the mean values of the $\Phi_{n}$ at
$m/\Lambda\to0$:
\begin{equation}
\langle\Phi_{n}\rangle \simeq C_{n}\nu_0^{-n} (m/\Lambda)^{\Delta_{n}}.
\label{9}
\end{equation}

Relation (\ref{9}) gives a contribution in $\langle\Phi_{n}\rangle$ that is
nonanalytic in $m/\Lambda$; this contribution prevails if $\Delta_{n}<0$
(the operator is ``dangerous''), which is indeed the case: in the first
order in $\varepsilon$ one obtains $\Delta_{n}=-2n(n-1)\varepsilon/(d+2)$,
cf. \cite{GK}. Strictly speaking, because of the mixing in the
renormalization
the monomial $\Phi_{n}$ becomes a finite linear combination of the
``scaling'' operators with definite dimensions $\Delta_{k}$ with $k\le n$;
in representations (\ref{7}) and (\ref{9}) we show only the leading term
with the maximum $|\Delta_{k}|$.

The scaling functions $F_{n}(mr)$, $F_{nk}(mr)$ in (\ref{6}), (\ref{7})
can be calculated as series in $\varepsilon$, but such
representations become useless if the behaviour of these functions at
$mr\to0$ is studied: the actual expansion parameter then becomes
$\varepsilon\ln(mr)$ rather than $\varepsilon$. Summation of such ``large
IR logarithms,''  in contrast with the ``large UV logarithms'' with
$\ln(\Lambda r)$, lies beyond the scope of the RG method in the narrow
sense of the word, because the forms of the scaling functions are not
determined by the plain RG equations. Like in the theory of critical
behaviour, the behaviour at $mr\to0$ can be extracted from the general
solution of the RG equations by means of the operator product expansion
(OPE). According to the OPE, the product of two (renormalized) composite
operators $O_{1}(x_{1})O_{2}(x_{2})$ (for example, $\theta^{n}$ or
renormalized analogues of the monomials $\Phi_{n}$) at
$t_{1}=t_{2}=t$, ${\bf r}\equiv {\bf x}_{1} - {\bf x}_{2}\to 0$
and the fixed ``centre of mass''
${\bf x}\equiv ({\bf x}_{1} + {\bf x}_{2} )/2 = {\rm const}$
is represented in the form
\begin{equation}
O_{1}(x_{1})O_{2}(x_{2})=\sum_{\alpha}C_{\alpha} ({\bf r})
O_{\alpha}({\bf x},t) ,
\label{SDE}
\end{equation}
where $C_{\alpha}$ are coefficients analytic in $(mr)^{2}$ and the
summation, in general case, runs over all local composite operators
$O_{\alpha}(x)$ with definite critical dimensions $\Delta_{\alpha}$.
Correlation functions like (\ref{6}), (\ref{7}) are obtained by
averaging relations (\ref{SDE}); the mean values $\langle O_{\alpha}(x)
\rangle\propto m^{\Delta_{\alpha}}$ appear on the right-hand sides,
they give rise to the contributions of the form $(mr)^{\Delta_{\alpha}}$
in the scaling functions $F_{n}(mr)$, $F_{nk}(mr)$. The leading terms at
$mr\to0$ are determined by the operator with the minimum possible value of
$\Delta_{\alpha}$. In conventional models like $\phi^{4}$ the leading term
is related to the simplest operator $O_{\alpha}(x)=1$ with
$\Delta_{\alpha}=0$. However, in model (\ref{1})--(\ref{3}) the critical
dimensions of all operators $\Phi_{n}$ are negative (see above), and they
determine the leading terms for $mr\to0$. Although the operators containing
fields $\theta$ without the derivative $\partial$ also have negative
dimensions, they do not contribute to the OPE for the quantities like
(\ref{6}), (\ref{7}) due to the invariance of the latter with respect to
the shift $\theta\to\theta+{\rm const}$ of the field $\theta(x)$.

If the operator product expansion for any given function (\ref{6}) or
(\ref{7}) included con\-t\-ri\-bu\-ti\-ons from all the operators $\Phi_{n}$,
we would have to sum all of them in order to obtain the behaviour for
$mr\to0$, because the spectrum of their dimensions is not bounded from below
(the most dangerous operator does not exist). This is the case in more
realistic models of fully developed turbulence; see [18--20]. Fortunately,
this problem does not exist in model (\ref{1})--(\ref{3}) due to the
linearity of equation (\ref{1}) in $\theta(x)$: one can show that the number
of fields $\theta$ in any composite operator $O_{\alpha}(x)$ that enters
the right-hand side of representation (\ref{SDE}) cannot exceed the total
number of fields on the left-hand side. We thus arrive at the following
expressions for quantities (\ref{6}), (\ref{7}):
\begin{equation}
S_{n}(r)\simeq \nu_0^{-n} r^{2n}(\Lambda r)^{-n\varepsilon}
\sum_{s\le n} C_{n,s} (mr)^{\Delta_{s}},
\label{16}
\end{equation}
\begin{equation}
S_{nk}(r)\simeq \nu_0^{-n-k} (\Lambda r)^{-\Delta_{n}-\Delta_{k}}
\sum_{s\le n+k} C_{nk,s} (mr)^{\Delta_{s}}
\label{17}
\end{equation}
with coefficients $C_{n,s}$, $C_{nk,s}$ dependent on $\varepsilon$ and $d$
and corrections of the form $(mr)^{2+O(\varepsilon)}$. Since the dimensions
$\Delta_{s}$ are negative and decrease as $s$ increases, the leading
asymptotic term in (\ref{16}), (\ref{17}) is given by the contribution with
the maximum $s$, that is, with $s=n$ in (\ref{16}) and $s=n+k$ in (\ref{17}):
\begin{equation}
S_{n}(r)\simeq C_{n}\nu_0^{-n} r^{2n}(\Lambda r)^{-n\varepsilon}
(mr)^{\Delta_{n}},
\label{11}
\end{equation}
\begin{equation}
S_{nk}(r)\simeq C_{nk}\nu_0^{-n-k} (\Lambda r)^{-\Delta_{n}-\Delta_{k}}
(mr)^{\Delta_{n+k}}.
\label{12}
\end{equation}

The critical dimensions are calculated as series of the form
$\Delta_{n} = \sum_{k=1}^{\infty} \varepsilon^{k} \Delta_{n}^{(k)}$;
the first-order term was already given above. We have calculated the
dimensions $\Delta_{n}$ in the second \cite{RG} and third \cite{cube,kub}
orders of the $\varepsilon$ expansion. The second-order result has the form:
\begin{eqnarray}
\Delta_n^{(2)} = \frac{n(n-1)}{(d-1)(d+2)^{3}(d+4)^{2}}\,
\Bigl[-4(d+1)(d+4)^{2}+
\nonumber \\
+ 3(d-1)(d+2)(d+4)(d+2n)h(d)- 4(d+1)(d+2)(d+3n-2)h(d+2) \Bigr] ,
\label{13}
\end{eqnarray}
where $h(d)\equiv F(1,1;d/2+2;1/4)$ and $F(\cdots)$ is the hypergeometric
series. Simpler expressions are obtained for integer $d$, in particular,
$h(2) = 8[1-3\ln(4/3)]$, $h(3) = 10 (\pi\sqrt 3-16/3)$, while for the other
integer $d$ analogous expressions can be obtained from the recurrent
relation $3h(d)+(d+2)h(d+2)/(d+4)=4$. The third-order coefficient is
presented in \cite{cube,kub}.

From the above formulas one obtains $\Delta_{1}^{(2)}=0$ in agreement with
the exact result $\Delta_{1}=0$. The formal proof of the latter is based on
certain Schwinger equation, which has the meaning of the conservation law
for the ``energy'' $\theta^{2}(x)$; see \cite{RG}. This means that the
second-order structure function $S_{2}$ is not anomalous in agreement with
the well-known exact solution obtained in \cite{Kraich1}.

We note that the family of operators $\Phi_{n}$ is ``closed with respect to
the fusion'' in the sense that the leading term of the correlation function
$\langle \Phi_{n} \Phi_{m} \rangle$ is given by the operator $\Phi_{n+m}$
from the same family with the summed index $n+m$. This fact along with the
inequality $\Delta_{n}+\Delta_{m} >  \Delta_{n+m}$, which is obvious from
the explicit expressions for  $\Delta_{n}$, can be interpreted as the
statement that the correlations of the local dissipation rate in the model
(\ref{1})--(\ref{3}) exhibit multifractal behaviour, cf. \cite{DL,Ey}.

An important issue that can be discussed on the example of the rapid-change
model is that of the nature and convergence of $\varepsilon$ series in
models of turbulence and the possibility of their extrapolation to finite
values of $\varepsilon\sim1$. The knowledge of the three terms of the
$\varepsilon$ expansion in model (\ref{1})--(\ref{3}) allows one to discuss
its convergence properties and to obtain improved predictions for finite
$\varepsilon$ in reasonable agreement with the existing nonperturbative
results: analytical and numerical solutions of the zero-mode equations
\cite{Falk1,Pumir} and numerical experiments \cite{VMF,MM}.

The RG and OPE approach presented above can be generalized to the cases
where compressibility, anisotropy or finite correlation time are present
[10--14] and the passive advection of vector (e.g., magnetic) fields
[15--17].

Let us conclude with a brief discussion of the RG and OPE approach to a more
realistic model of fluid turbulence: the stirred Navier--Stokes equation;
see [18--20]. Dangerous operators in that model are absent in the
$\varepsilon$ expansions and can appear only at finite values of
$\varepsilon$. This means that they can be reliably identified only if
their dimensions are derived {\it exactly} with the aid of Schwinger
equations or Galilean symmetry. Due to the nonlinear nature of the problem,
they enter the corresponding OPE's as
infinite families whose spectra of dimensions are not bounded from below,
and in order to find the small-$mr$ behavior one has to sum up all their
contributions in the representations like (\ref{11}), (\ref{12}). The
needed summation of the most singular contributions, related to the operators
of the form $v^{n}$ with known dimensions, was performed in \cite{JETP}
using the so-called infrared perturbation theory for the case of the
different-time pair correlation functions. It has revealed their strong
dependence on $m$, which physically can be explained by the well-known
``sweeping effects.'' This demonstrates that, contrary to the existing
opinion \cite{CK,Woo}, the sweeping effects can be properly described
within the RG approach, but one should combine the RG and OPE techniques
and go beyond the plain $\varepsilon$ expansions. Analysis of the $m$
dependence of the Galilean invaiant objects like the structure functions
requires the explicit construction of all dangerous invariant scalar
operators, exact calculation of their critical dimensions, and summation
of their contributions in the corresponding OPE. This is clearly not a
simple problem and it requires considerable improvement of the present
techniques.

The work was supported by the Nordic Grant for Network Cooperation with the
Baltic Countries and Northwest Russia No.~FIN-18/2001 and the GRACENAS Grant
No.~E00-3-24.

\end{document}